\documentstyle[prl,aps,twocolumn,psfig]{revtex}
\begin{document}
\draft
\title{Fermi-level alignment at metal-carbon nanotube interfaces: 
application to scanning tunneling spectroscopy}
\author{Yongqiang Xue \cite{yxue} and Supriyo Datta}
\address{School of Electrical and Computer Engineering, Purdue 
University, West Lafayette, IN 47907}
\date{\today}
\maketitle
\begin{abstract}
At any metal-carbon nanotube interface there is charge transfer and the 
induced interfacial field determines the position of the carbon nanotube 
band structure relative to the metal Fermi-level. In the case of a 
single-wall carbon nanotube (SWNT) supported on a gold substrate, we show 
that the charge transfers induce a local electrostatic potential 
perturbation which gives rise to the observed Fermi-level shift in scanning 
tunneling spectroscopy (STS) measurements. We also discuss the relevance of 
this study to recent experiments on carbon nanotube transistors and argue 
that the Fermi-level alignment will be different for carbon nanotube 
transistors with low resistance and high resistance contacts.
\end{abstract}
The discovery of carbon nanotube opens up a new artificial laboratory in 
which one-dimensional transport can be investigated\cite{Dekker}, similar 
to the semiconductor quantum wire\cite{Datta1}. However, the study of 
transport in carbon nanotube has been complicated by the difficulty of 
making low resistance contacts to the measuring 
electrodes. The high resistances reported in various two- and three-terminal 
measurements \cite{Dekker1} have led Tersoff \cite{Tersoff1} (and 
also the present authors \cite{Datta2}) to suggest that 
wavevector conservation at the metal-carbon nanotube contact may play 
an important role in explaining the high contact resistance. 
In this paper we address a different question: How does the Fermi-level in 
the metallic contact align with the energy levels of the nanotube? The 
answer to this question is very important in interpreting the transport 
measurements. Depending on the contact geometry, 
transport can occur in the direction parallel to the nanotube axis, in the 
case of nanotube field-effect-transistor (FET) \cite{Dekker1,Dai1}, or 
perpendicular to it, in the case of the STS measurement\cite{Dekker2}. 
In the STS measurement, the Fermi-level is found to have shifted to the 
valence band edge of the semiconducting nanotube\cite{Dekker2}, which is 
then used to explain the operation of the nanotube FETs with high 
resistance contacts\cite{Dekker1}, where the measured two-terminal 
resistance for metallic nanotube is $\sim 1M\Omega$. However, low 
temperature transport measurements using low resistance 
contacts\cite{Dai1} (where the 
contact resistance is of the order of resistance quantum) indicate that the 
Fermi-level is located between the valence and conduction band of the 
semiconducting nanotube, instead of 
being pinned to the valence band edge. This conflict raises the 
important question of whether the Fermi-level positioning may depend on the 
contact geometry and/or the interface coupling.

In this paper we present a theory of the scanning tunneling spectroscopy 
of a single-wall carbon nanotube (SWNT) supported on the Au(111) substrate. 
The main results of our work are: {\bf (1)} the work function difference 
between the gold substrate and the nanotube leads to charge transfers across 
the interface, which induces a local electrostatic potential perturbation on 
the nanotube side giving rise to the observed Fermi-level shift in the 
STS measurement. 
{\bf (2)} for nanotube transistors, the atomic-scale 
potential perturbation at the interface is not important 
\emph{if the coupling between the metal and the 
nanotube is strong}. The metal-induced gap states (MIGS) model 
provides a good starting point for determining the Fermi-level position. 
{\bf (3)} a proper theory of STS should take the tip electronic structure 
into account. 

For an ordinary metal-semiconductor interface, the MIGS model provides a 
conceptually simple way of understanding the band lineup problem which 
predicts that the metal Fermi-level $E_{F}$ should align with the ``charge 
neutrality level'' (which can be taken as the energy where the gap states 
cross over from valence- to conduction-type character) in the 
semiconductor\cite{Tersoff2}. This elegant idea has been applied with 
impressive success by Tersoff to various metal-semiconductor 
junctions and semiconductor heterojunctions, which greatly simplifies 
the band lineup problem and gives quantitatively accurate prediction of the 
Schottky barrier height in many cases\cite{Tersoff2}. 
The success of this model relies on the fact that there exists a continuum of 
gap states around $E_{F}$ at the semiconductor side of the metal-semiconductor 
interface due to the tails of the metal wavefunction decaying into the 
semiconductor, which can have significant amplitude over a few atomic 
layers near the interface\cite{Heine}. Any deviation from local 
charge neutrality in the interface region will result in metallic screening 
by the MIGS. 

\begin{figure}
\centerline{\psfig{figure=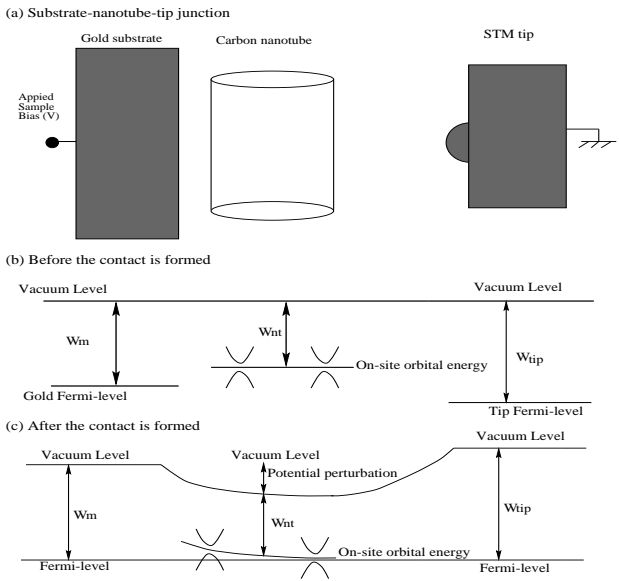,angle=0.,height=3.0in,width=3.2in}}
\vspace{0.1cm} \caption{Illus\-tra\-tion of the formation of a 
substrate-SWNT contact. We only show a semiconducting SWNT here. 
The work functions of the gold, the SWNT and the platinum tip are 
$W_{m}=5.3 (eV)$, $W_{nt}=4.5 (eV)$ and $W_{tip}=5.7 (eV)$ respectively. 
(c) shows our picture of the interface Fermi-level positioning. }
\label{xueFig1}
\end{figure}

However, this is not true for the interface formed when a SWNT is deposited onto 
the gold substrate. Since the coupling to the substrate is weak and the 
metal wave function decays across a significant van der Waals 
separation\cite{Tersoff1,Avouris}, the MIGS will provide only relatively 
weak screening. When the conductance spectrum is measured using a 
scanning tunneling microscope (STM), transport occurs perpendicular to the 
nanotube axis and the characteristic length scale is the diameter of the SWNT 
which is on the scale of nanometers\cite{Dekker2} and can be comparable to the 
range of the interfacial perturbation. The detailed potential variations 
in this dimension will be important in determining the STS current-voltage 
characteristics, similar to the case of molecular adsorbates on metal 
surfaces\cite{Xue}. Fig.\ \ref{xueFig1} illustrates schematically 
the local electrostatic potential profile at the substrate-nanotube-tip 
heterojunction. If the charge distributions on both sides don't change when 
interface is formed, then the vacuum levels line up\cite{Tersoff2}. 
However, due to the difference of work functions\cite{Dekker1,Dekker2} (as 
shown in Fig.\ \ref{xueFig1}(b)), electrons will transfer from the SWNT to 
the gold substrate and the resulting electrostatic potential 
profile $\delta \phi$ should be determined self-consistently (Since the 
perturbation due to the tip is much weaker, we neglect its effect when 
treating the substrate-SWNT interface). 

We assume an ideal substrate-SWNT interface and study the interface 
electronic structure using the $\pi$-electron tight-binding (TB) model of 
the SWNT\cite{Dress1}. In this model, 
the bandstructure of SWNTs is symmetric with respect to the position of 
the on-site  $\pi$ orbital energy. We take the Fermi-level of the gold as the 
energy reference, then the initial $\pi$ orbital energy at each carbon atom of 
the SWNT is $W_{m}-W_{nt}=0.8 (eV)$. The final on-site $\pi$ orbital energy 
is the superposition of this initial value and the change in the electrostatic 
potential $\delta \phi$ which changes as one moves away from the gold 
substrate (Fig.\ \ref{xueFig1}(c)). For the gold substrate we use the TB 
parameters of Papaconstantopoulos \cite{Papa}. For the coupling between the 
SWNT and the gold surface, we use the values obtained from the Extended 
H\"{u}ckel Theory (EHT) \cite{Hoffman}. Only the carbon atoms closest to the 
gold surface are assumed to be coupled\cite{Note1}. 

\begin{figure}
\centerline{\psfig{figure=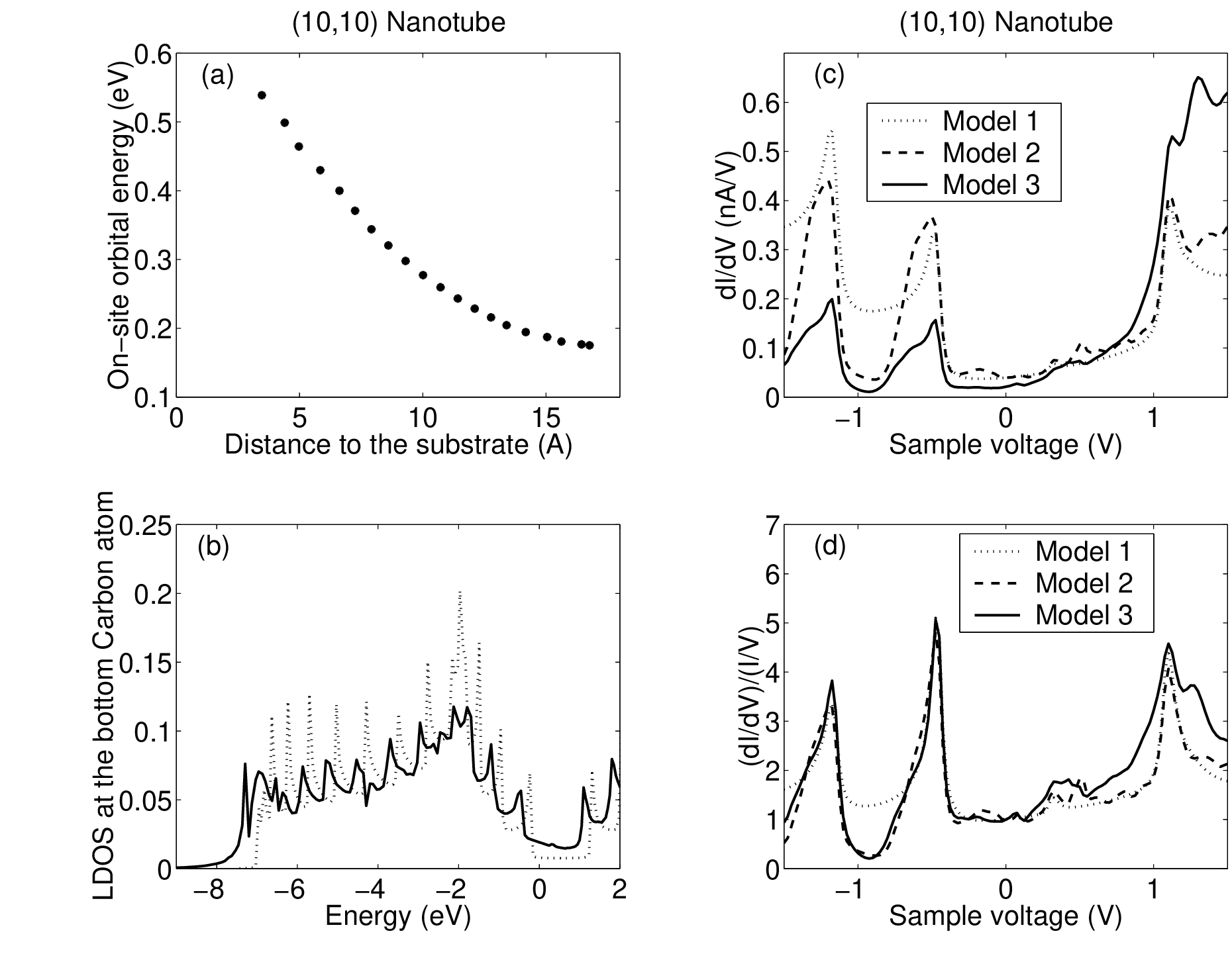,angle=0.,height=3.1in,width=3.2in}}
\vspace{0.1cm} \caption{Calculated results for (10,10) nanotube. (b) shows 
the LDOS at the carbon atom closest to the substrate in unit of 
states/$eV$/atom, the dotted line is that 
of the isolated SWNT horizontally shifted by the corresponding 
potential perturbation $\delta \phi_{i}$. (c) and (d) show the calculated 
STS $dI/dV - V$ and $d\ln I/d\ln V - V $ curves. Model 1: tip modeled as 
having a constant density of states $\rho_{tip}$; Model 2: tip modeled as a 
semi-infinite Pt(111) crystal; Model 3: tip modeled as a Pt atom adsorbed 
on the surface of the semi-infinite Pt(111) crystal.  } 
\label{xueFig2}
\end{figure}

Since the SWNT has periodic symmetry along its axis, only one unit 
cell needs to be considered. We use the Green's function method to calculate 
the electron population of each carbon atom from the expression:
 $n_{i}=-\frac{2}{\pi} Imag\{ \int_{-\infty}^{E_{F}} G_{i,i}(E)dE \}$
where $G(E)$ is the projection of the Green's function onto one unit cell 
of the SWNT and $G_{i,i}$ is the $i$th diagonal matrix element 
corresponding to atom $i$ in the unit cell. $G(E)$ is calculated by reducing 
the Hamiltonian of the whole interface into an effective one in which the 
interactions between the given unit cell and the rest of the interface system 
are incorporated into the corresponding self-energy operators using the same 
method as described in ch.\ 3 of Datta \cite{Datta1}. Within the 
tight-binding theory, self-consistency is achieved by adjusting the diagonal 
elements of the Hamiltonian and imposing Hartree consistency 
between the potential perturbation $\delta \phi_{i}$ and the charge 
perturbation $\delta n_{i}$ using the self-consistent scheme similar to 
that developed by Flores and coworkers \cite{Flores} and also 
Harrison \cite{Harrison} (for details see Ref.\ \cite{Xue2}).
  
\begin{figure}
\centerline{\psfig{figure=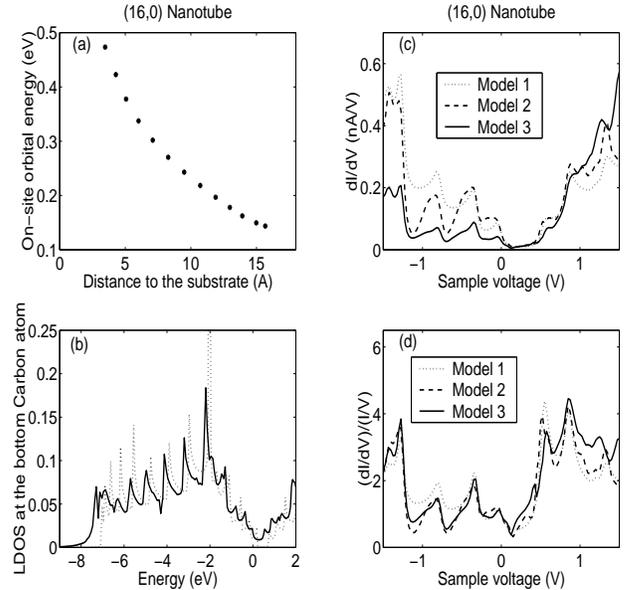,angle=0.,height=3.1in,width=3.2in}}
\vspace{0.1cm} \caption{ Calculated results for (16,0) nanotube, otherwise 
same as Fig.\ \ref{xueFig2}.}\label{xueFig3}
\end{figure}

Fig.\ \ref{xueFig2}(a)-(b) and Fig.\ \ref{xueFig3}(a)-(b) show the results 
for (10,10) and (16,0) SWNTs with diameters of $1.35$ and $1.25 (nm)$ 
respectively, close to those measured in Ref.\ \cite{Dekker2}. The 
substrate-SWNT distance is $3.2 (\AA)$\cite{Note}. We have 
also studied (15,0) and (14,0) SWNTs. All nanotubes show similar 
behavior. We believe similar conclusions can be reached for chiral 
nanotubes since the electronic structure of SWNTs depends only on their 
metallicity and diameter, not on chirality\cite{M&W}. The similarity 
between the metallic and the semiconducting nanotube shown here can be 
understood from the work of Benedict et al.\ \cite{Louie}, who show that the 
dielectric response of SWNTs in the direction perpendicular to the axis 
doesn't depend on the metallicity, only on the diameter. 

Since the $\pi$ orbital energy coincides with the position of the Fermi-level 
(mid-gap level) of the isolated metallic (semiconducting) SWNT, then the 
Fermi-level shift in the STS measurement should correspond to the on-site 
$\pi$ orbital energy of the carbon atom closest to the STM tip \emph{if 
only this atom is coupled to the tip}. However, considering the cylindrical 
shape of the SWNT, more carbon atoms could be coupled to the tip and the 
Fermi-level shift then corresponds to the average value of the on-site orbital 
energies of the carbon atoms within the coupling range. From the plotted 
values of Fig.\ \ref{xueFig2} and Fig.\ \ref{xueFig3}, we then expect 
Fermi-level shifts of $\sim 0.2 (eV)$ for both nanotubes, close to 
the measured values\cite{Dekker2}. The peak structures in the local 
density-of-states (LDOS) of the bottom carbon atom (closest to the gold 
substrate) corresponding to the Van Hove singularities are broadened due to 
the hybridization with the gold surface atomic orbitals. Their positions also 
change, which can be understood from the bonding-antibonding splitting 
resulting from the hybridization of the nanotube molecular orbitals and 
the gold orbitals. Also notable is the enhancement of the density of states 
in the gap at the expense of the valence band, reminding us of the Levinson 
theorem which states the total number of states should be conserved in the 
presence of perturbation, be it due to impurity or due to surface\cite{Appel}. 
In contrast, the perturbation of the LDOS at the carbon atom furthest to 
the substrate is much weaker. The calculated 
charge transfer per atom is small and mainly localized 
on the carbon atoms close to the gold surface, in agreement with recent 
\emph{ab initio} calculations\cite{Rubio}.

\emph{Applications to the scanning tunneling spectroscopy.}
The differential 
conductance $dI/dV$ (or the normalized one $d\ln I/d\ln V$) obtained from 
the STS measurement is often interpreted as to reflect the local density of 
states of the sample, based on the s-wave model of the tip\cite{Tersoff3}. 
However, first-principles calculations have shown this model to be inadequate 
for tips made from transition metals, where small clusters tend to form at 
the tip surface giving rise to localized d-type tip states\cite{W&G}. As a 
result, the tip electronic structure can 
have profound effects on the interpretation of the STS 
measurement\cite{Xue,W&G}. 

The STS current-voltage characteristics can be calculated using the standard 
technique of scattering theory\cite{Datta1,Xue}. Here we have taken a 
simpler approach instead, aiming only to illustrate how the tip electronic 
structure may affect the interpretation of the STS measurement. Since the 
coupling across the SWNT-tip interface is weak, the 
tunneling Hamiltonian theory may be invoked to write the current crudely as:
\begin{equation}
 I \propto \int_{0}^{eV} \rho_{nt}(E)\rho_{tip}(E-eV)dE
\label{Tunneling}
\end{equation}
where $\rho_{nt}$ and $\rho_{tip}$ are the density of states of 
the SWNT and the tip respectively. The differential 
conductance thus obtained then reflects the convolution of the density 
of states of the SWNT and the tip. If $\rho_{tip}$ is constant within the 
range of the integral, we 
recover the usual expression $dI/dV \propto \rho_{nt}$. Note $\rho_{nt}$ is 
calculated taking the on-site perturbations and the coupling to 
the gold substrate into account (we use the LDOS of the carbon atom closest 
to the tip here). We have used two models for the tip: (1) as 
a semi-infinite Pt(111) crystal; (2) as a Pt atom 
adsorbed on the surface of the semi-infinite Pt(111) crystal \cite{Xue}. The 
results are shown in Fig.\ \ref{xueFig2}(c)-(d) and Fig.\ 
\ref{xueFig3}(c)-(d) along with that obtained from Eq.\ (\ref{Tunneling}) 
assuming constant $\rho_{tip}$. As can be seen from the plots, 
additional fine structures are introduced between the peak structures of 
$\rho_{nt}$ when we take into account the electronic structure of the tip.

\emph{Discussions and conclusions} 
With the advancement of new techniques for making electric contact to the 
SWNT \cite{Dai1,Dai2}, low resistance contacts with two-terminal 
conductance close to the conductance quantum have been obtained \cite{Ron}.  
Current-voltage characteristics measured at low temperature using these new 
techniques show that the Fermi-level is located in the gap of the 
semiconducting SWNT. In these experiments, SWNTs are 
grown from the patterned catalyst islands on the silicon wafer, Au/Ti contact 
pads are then placed on the catalyst islands fully covering the islands and 
extending over their edge\cite{Dai1}. Since the SWNTs 
thus grown are mostly capped\cite{Dai2}, the coupling between the SWNTs and 
the electrode is presumably similar to that of fullerene where it is well 
known that fullerene forms a strong chemical bond with the noble and 
transition metal surfaces (see Dresselhaus et al.\ \cite{Dress2}). The large contact 
area between the SWNTs and the metal will make the coupling across the 
interface even stronger which then allows the metal wavefunctions to 
penetrate deep into the nanotube side. 
Therefore, we expect that the dominant contribution to the barrier height 
is from the metallic screening by MIGS, which tends to line up the metal 
Fermi-level with the ``charge neutrality level'' of the SWNT. Since the band 
structure of the SWNT is exactly symmetric within the 
$\pi$-electron model, the ``charge neutrality level'' will be at the 
mid-gap, although it can be different when a more accurate model of 
electronic structure is used. Our emphasis here is not to give a quantitative 
estimate of the barrier height, but rather to show that the MIGS model 
provides the conceptual base of understanding in the limit of strong 
interface coupling. The situation gets 
complicated for measurements using high resistance contacts, where the SWNT 
is side-contacted \cite{Dekker1} and the coupling across the interface is 
weak\cite{Tersoff1,Avouris}. In this case, the MIGS model of Schottky 
barrier is no longer applicable. 
The interface defects and bending of SWNT at the edge of the contact 
can induce localized states at the interface region\cite{Avouris} which will 
accommodate additional charges and affect the formation of Schottky 
barrier\cite{Monch}. Therefore, we expect that the final Fermi-level 
position depends on the detailed contact condition and \emph{may or 
may not be located at the valence band edge}. 
We believe that a detailed \emph{ab initio} analysis is needed to clarify the 
various mechanisms involved. 

This work is jointly supported by NSF and ARO through grant number 
9708107-DMR. We are indebted to M.P. Anantram for drawing our attention 
to this important topic.


\begin{references}
\bibitem{Note} At this distance, the coupling across the interface is weaker 
than the bulk correspondents. This is appropriate since the coupling between 
the SWNT and the substrate is not strong chemical bonding. See T. Hertel, 
R. Martel and Ph. Avouris, J.\ Phys.\ Chem.\ B {\bf 102}, 910 (1998).
\end{references}
\end{document}